\documentclass[a4paper,11pt]{article}
\usepackage{pos, pos_e, packets}
\usepackage[capitalize]{cleveref}
\graphicspath{{figures/}}

\title{Machine-Learning-Enhanced Optimization of Noise-Resilient Variational Quantum Eigensolvers}
\ShortTitle{Machine-Learning-Enhanced Optimization of Noise-Resilient VQEs}

\author*[a,b]{Kim A.\ Nicoli}
\author*[a,b]{Luca J.\ Wagner}
\author[a,b]{Lena Funcke}

\affiliation[a]{Transdisciplinary Research Area ``Building Blocks of Matter and Fundamental Interactions'' (TRA Matter), University of Bonn, Regina-Pacis-Weg 3, 53113 Bonn, Germany}

\affiliation[b]{Helmholtz Institute for Radiation and Nuclear Physics (HISKP), University of Bonn, Nussallee 14-16, 53115 Bonn, Germany}

\emailAdd{knicoli@uni-bonn.de}
\emailAdd{lwagner@hiskp.uni-bonn.de}
\emailAdd{lfuncke@uni-bonn.de}

\abstract{

Variational Quantum Eigensolvers (VQEs) are a powerful class of hybrid quantum-classical algorithms designed to approximate the ground state of a quantum system described by its Hamiltonian. VQEs hold promise for various applications, including lattice field theory.
However, the inherent noise of Noisy Intermediate-Scale Quantum (NISQ) devices poses a significant challenge for running VQEs as these algorithms are particularly susceptible to noise, e.g., measurement shot noise and hardware noise.

In a recent work, it was proposed to enhance the classical optimization of VQEs with Gaussian Processes (GPs) and Bayesian Optimization, as these machine-learning techniques are well-suited for handling noisy data.
In these proceedings, we provide additional insights into this new algorithm and present further numerical experiments. In particular, we examine the impact of hardware noise and error mitigation on the algorithm's performance. 
We validate the algorithm using classical simulations of quantum hardware, including hardware noise benchmarks, which have not been considered in previous works.
Our numerical experiments demonstrate that GP-enhanced algorithms can outperform state-of-the-art baselines, laying the foundation for future research on deploying these techniques to real quantum hardware and lattice field theory setups.

}

\FullConference{%
  The 41st International Symposium on Lattice Field Theory (Lattice2024)\
  29 July-02 August 2024\
  Liverpool, United Kingdom
}

\begin{document}
\maketitle

\section{Introduction}
\label{sec:Introduction}

Quantum hardware with several hundreds of qubits can already be harnessed to outperform classical computers on specific tasks (see, e.g., Ref.~\cite{Morvan:2023inh}). These tasks, however, currently have no practical applications and are specifically designed to be challenging for classical computers. 
As we have entered the Noisy Intermediate-Scale Quantum (NISQ) era of quantum computing, the primary challenge is to develop algorithms for NISQ devices that may demonstrate quantum advantage in tasks of practical relevance, such as quantum chemistry~\cite{mccaskey2019quantumchemistrybenchmarknearterm} or quantum field theories~\cite{DiMeglio:2023nsa,Funcke:2023jbq}.

Particularly relevant for NISQ devices are hybrid quantum-classical algorithms, such as the Variational Quantum Eigensolver (VQE)~\cite{Peruzzo_2014, McClean_2016}.
VQEs can approximate the ground state of quantum Hamiltonians by classical variational optimization of a parametrized quantum circuit. The quantum computer measures the energy of the quantum system for a given set of parameters, i.e., the objective of the minimization problem, while a classical optimization routine is used to find a better set of parameters. Efficient exploration of the parameter space is crucial for successful optimization. 

Nakanishi et al.~\cite{nakanishi20} demonstrated that, under certain conditions, the functional form of the VQE's objective function can be explicitly derived and utilized to develop a more efficient optimization routine. The resulting Nakanishi-Fuji-Todo (NFT) algorithm~\cite{nakanishi20} uses two measurements to \emph{sequentially} optimize one parameter of the quantum circuit at a time, by fitting the objective in a one-dimensional subspace. Indeed, this constitutes a sequential minimal optimization (SMO)~\cite{Platt1998SequentialMO}, which differs from global optimization protocols such as gradient-based approaches. 

Despite its efficiency, the NFT algorithm faces several challenges, primarily due to the noise of current quantum hardware. In particular, the algorithm relies on noisy measurements to identify the global minimum in the one-dimensional subspace, which often hinders convergence and complicates optimization. 
Machine learning offers a promising approach to addressing the challenge of noisy measurements and enhancing the NFT algorithm with more flexibility. In a recent work~\cite{nicoli2023physicsinformed}, Nicoli et al. proposed an optimization algorithm called NFT-with-EMICoRe (EMICoRe for short), which extends the NFT framework by incorporating Gaussian Process Regression (GPR) and Bayesian Optimization (BO) for a data-driven and potentially more noise-resilient SMO scheme. However, this work~\cite{nicoli2023physicsinformed} focused solely on measurement shot noise, neglecting the impact of quantum hardware noise, such as decoherence, depolarization, and crosstalk errors.

In these proceedings, we build on prior works~\cite{nakanishi20,nicoli2023physicsinformed} and demonstrate that EMICoRe outperforms the NFT baseline in the presence of simulated quantum hardware noise. We start in Sec.~\ref{sec:Pre} by reviewing the theoretical foundations of VQE, Gaussian Processes, Bayesian Optimization. In Sec.~\ref{sec:ProposedMethod}, we introduce the EMICoRe algorithm. This is followed by numerical experiments in Sec.~\ref{sec:experiments}, which involve classical simulations of noisy quantum hardware, comparing the results with and without error mitigation. We conclude and provide an outlook in Sec.~\ref{ref:conclusion}.

\section{Theoretical Foundations}
\label{sec:Pre}

\subsection{Variational Quantum Eigensolver (VQE)}
\label{sec:VQE}

Computing the ground state energy of a quantum system described by a Hamiltonian $H$ and the corresponding ground state wave function $\ket{\psi_\mathrm{GS}}$ is a key challenge in many fields of physics.
The Hamiltonian of an $N$-qubit system can be represented as
\begin{equation}\label{eq:Pauli_Hamiltonian}
H=\sum_{\alpha=1}^T h_\alpha P_\alpha ,
\end{equation}

where $P_\alpha \in \{X, Y, Z, I\}^{\otimes N}$ is a tensor product of single-qubit Pauli operators $X, Y, Z$ and the identity operator $I$. The $h_\alpha$ are real coefficients, and $T$ is the number of individual Pauli terms.

In general, it is challenging to compute the ground state wave function $\ket{\psi_\mathrm{GS}}$ because it requires diagonalizing the Hamiltonian $H$, which is a $\mathbb{C}^{2^N \times 2^N}$-dimensional matrix for an $N$-qubit system and thus scales exponentially with the number of qubits. An alternative approach to exact diagonalization is to compute a variational approximation of the ground state. This can be achieved, for instance, by using the Variational Quantum Eigensolver (VQE)~\cite{Peruzzo_2014, McClean_2016}, where a parametric quantum state $|\psi(\boldsymbol{\theta})\rangle$ is prepared on a quantum computer. This quantum state, which depends on $D$ angular parameters $\boldsymbol{\theta} \in [ 0, 2\pi)^D$, is obtained through a sequence of $D'$ ($\geq D$) unitary quantum gate operations $U(\boldsymbol{\theta}) = U_1 \cdots U_{D'}$ acting on an initial state $|\psi_0\rangle$.  A subset $D$ of these operations are Pauli rotation gates,  $U_{d'(d)}(\theta_d) = R_{P_d}(\theta_d) = \exp(-i \theta_d P_d)$, where $P_d \in \{ X, Y, Z \}$ are the Pauli matrices and $\theta_d$ are the angular parameters. The full variational state can thus be written as $|\psi(\boldsymbol{\theta})\rangle = U(\boldsymbol{\theta}) |\psi_0\rangle$. After the preparation of the quantum state $|\psi(\boldsymbol{\theta})\rangle$ on the quantum computer, the state is measured in the basis of the individual Pauli strings to obtain the corresponding \textit{measured energy} given by
\begin{align}
\tilde{E}(\boldsymbol{\theta}) &=E^*(\boldsymbol{\theta})+\varepsilon, \quad \label{eq:energy}\\
\textrm{where} \quad E^*(\boldsymbol{\theta})&=\left\langle\psi(\boldsymbol{\theta})|H| \psi(\boldsymbol{\theta})\right\rangle=\left\langle\psi_0\left|U(\boldsymbol{\theta})^{\dagger} H U(\boldsymbol{\theta})\right| \psi_0\right\rangle \text {. }
\end{align}
The energy $E^\ast(\boldsymbol{\theta})$ is the \textit{true energy} that would be obtained if 
the quantum computer were noise-free. Typically, a quantum computer is subject to both Gaussian noise, arising from the statistical nature of repeated measurements (i.e., \textit{measurement shot noise}), and \textit{hardware noise}, which stems from imperfections in qubits, gates, and measurements~\cite{Georgopoulos_2021}. Both types of noise are encoded in the noise term $\varepsilon$ in~\cref{eq:energy}. 

\begin{figure}
    \centering
    \includegraphics[width=0.8\textwidth]{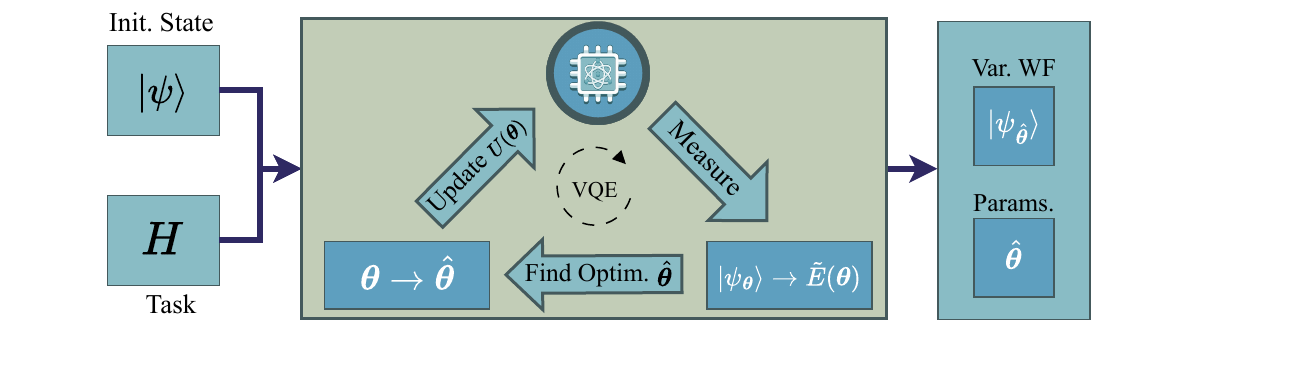}
    \caption[Procedure of a typical VQE]{Illustration of the VQE workflow. For more details, see the text.}
    \label{fig:VQE}
\end{figure}

As shown in the green box of~\cref{fig:VQE}, the VQE protocol can be divided into the following steps:
\begin{enumerate}
    \item Initialize a random quantum state with random parameters $\boldsymbol{\theta}$.
    \item \label{step:measure_VQE} Measure the expectation value $\Tilde{E}(\boldsymbol{\theta}) = \left\langle\psi(\boldsymbol{\theta})|H| \psi(\boldsymbol{\theta})\right\rangle + \varepsilon$ on the quantum computer.
    \item Use a classical non-linear optimizer, such as gradient-based or gradient-free methods, to determine new parameters ${\boldsymbol{\hat{\theta}}}$ that minimize $\tilde{E}(\boldsymbol{\theta})$.
    \item Iterate to step~\ref{step:measure_VQE} until  $\tilde{E}(\boldsymbol{\theta})$ converges to the ground state energy.
\end{enumerate}

Under mild assumptions for the above described variational circuit, Nakanishi et al.\ derived an analytical expression of the VQE's objective functional~\cite{nakanishi20},
\begin{equation}\label{eq:VQE_objective}
    E^\ast(\boldsymbol{\theta}) = \boldsymbol{b}^\top \cdot \left[\bigotimes^D_{d=1} \left(\begin{array}{c} \cos{(\theta_d)} \\\sin{(\theta_d)} \\ 1\end{array}\right)\right]\textrm{ , } \quad \forall\boldsymbol{\theta} \in [ 0, 2\pi)^D,
\end{equation}
where $\boldsymbol{b}$ is a set of arbitrary coefficients and $\theta_d$ is a $D$-dimensional vector of angular parameters for the variational quantum circuit.
In particular, this functional form leads to the following observation: when keeping all but one parameter constant, i.e., 
\begin{equation}
\boldsymbol{\theta}^{(t)}_{d} = (\theta_1, \ldots, \theta_{d-1}, \theta_d, \theta_{d+1}, \ldots, \theta_D)^\top \quad \mathrm{with} \quad \theta_{i\neq d} = \textrm{const.,}
\end{equation}
the form of the objective function for this one-dimensional subspace of the whole optimization problem reduces to~\cite{nakanishi20}
\begin{equation}\label{eq:VQE_objective_one_parameter}
\begin{aligned}
    E^{\ast(t)}(\boldsymbol{\theta}^{(t)}_d) &= \sum_{k=1}^K h_k\left\langle\psi_k\left|U_d^{ \dagger}\left(\theta^{(t)}_d\right) P_k U_d\left(\theta^{(t)}_d\right)\right| \psi_k\right\rangle \\
    &= a^{(t)}_{1, d} \cos \left(\theta^{(t)}_d-a^{(t)}_{2, d}\right)+a^{(t)}_{3, d},
\end{aligned}
\end{equation}
where $a^{(t)}_{\ell, d}(\ell=1,2,3)$ denote parameters that are independent of $\theta^{(t)}_d$, while the superscript $(t)$ refers to the optimization step. As a result, measuring three points in the one-dimensional subspace is sufficient to fit a function across the entire domain and analytically determine its minimum.

The NFT optimization algorithm~\cite{nakanishi20} uses this property and sequentially optimizes each parameter in the set $\boldsymbol{\theta}$ by requiring (at least) two measurements for each optimization step. The new observed points are \textit{deterministically} chosen along the axis $d$, i.e., 
\begin{align}\label{eq:shifts}
\boldsymbol{\Theta}^{\prime}=\left(\boldsymbol{\theta}_1^{\prime}, \boldsymbol{\theta}_2^{\prime}\right)=\left\{{\boldsymbol{\hat{\theta}}}^{(t-1)}- \alpha \boldsymbol{e}_d, {\boldsymbol{\hat{\theta}}}^{(t-1)}+ \alpha \boldsymbol{e}_d\right\}\,, 
\end{align}
where $\boldsymbol{\hat{\theta}}^{(t-1)}$ denotes the previous best point from step $(t-1)$.\footnote{The original paper from Nakanishi et al. \cite{nakanishi20} proposed a shift $\alpha$ of $\frac{\pi}{2}$ for~\cref{eq:shifts}. In the work by Nicoli et al. \cite{nicoli2023physicsinformed} it was suggested that equidistant points for probing the subspace should be used, i.e., a shift of $\frac{2\pi}{3}$ should be preferred. Later, it was derived by Anders et al.~\cite{andersadaptive} that the offset of $\frac{2\pi}{3}$ provides a constant variance over the whole 1D-subspace. In a private communication with the authors, it was reported that, because of a bug, the actual value used in the experiments of~\cite{nicoli2023physicsinformed} was $\frac{\pi}{3}$ instead of $\frac{2\pi}{3}$ as reported in the paper. Although this did not lead to significant differences in the numerical experiments, in order to be consistent with the prior work, we will stick to the convention of using $\frac{\pi}{3}$ for NFT.\label{footnote}} 
This approach is advantageous because it enables analytical optimization of each parameter within its respective one-dimensional subspace, eliminating the need to compute gradients as required by gradient-based methods, such as SPSA~\cite{SPSA_optimization_initial,SPSA_optimization_noisy_observation}. Furthermore, gradient-based approaches may optimize all parameters simultaneously during each gradient update but do not guarantee analytical minimization.

\subsection{Gaussian Processes and Bayesian Optimization}
\label{sec:Pre.GPandBO}

The NFT algorithm can find the \textit{exact} analytical minimum within the one-dimensional subspace if and only if the state preparation and measurements on the quantum computer are exact---that is, if an infinite number of shots are taken and no hardware noise is present. In a real-world scenario, due to the unavoidable hardware noise on current NISQ hardware and the inherent statistical uncertainty from repeated measurements (shots), it may be beneficial to enhance existing algorithms with frameworks like Bayesian Optimization (BO)~\cite{frazier2018tutorial} and Gaussian Process Regression (GPR)~\cite{rasmussen}.  These techniques, widely established in machine learning, are often well-suited to deal with noisy data~\cite{rasmussen}. Let $E^\ast(\cdot) : \mathcal{X} \mapsto \mathbb{R}$  be an unknown (black-box) objective function to be minimized (i.e., the energy in our case). With the aid of a function that suggests promising new points to observe, often referred to as \textit{acquisition function}~\cite{frazier2018tutorial}, one iteratively approximates the target objective with a surrogate function $E(\boldsymbol{\theta}) \approx E^\ast(\boldsymbol{\theta})$, sampled from the GP. At each optimization step, after new points are measured, those are added to the previous set of measured points, used to update the GP.

A common choice for a surrogate model is a GP regression model with a one-dimensional Gaussian likelihood and a GP prior~\cite{nicoli2023physicsinformed},
\begin{equation}\label{eq:GP_prior}
p(\tilde{E} \mid \boldsymbol{\theta}, E(\cdot))=\mathcal{N}_1\left(\tilde{E} ; E(\boldsymbol{\theta}), \sigma^2\right), \quad p(E(\cdot))=\operatorname{GP}(E(\cdot) ; \nu(\cdot), k(\cdot, \cdot)).
\end{equation}
Here, $\sigma^2$ is the variance of the noise $\varepsilon$ defined in~\cref{eq:energy}, and the prior mean function $\nu(\cdot)$ is set to $0$ in this work. In the case of VQEs, a zero-mean prior is a reasonable choice because the target function $E^\ast(\boldsymbol{\theta})$, which one aims to model, is a linear combination of functions centered around zero (up to a potential constant offset stemming from~\cref{eq:VQE_objective}). The function $k(\cdot, \cdot)$ is the kernel (covariance) function encoding the expected correlation between two measured points. We refer to ~\cref{sec:vqe_kernel} for more details on the kernel function we use in these proceedings.
The GP is trained on noisy measurements $\tilde{E} = E^\ast(\boldsymbol{\theta}) + \varepsilon$ that were collected throughout previous optimization steps. 

Let $\{\boldsymbol{\Theta}, \boldsymbol{\tilde{E}}\}$ be $N$ training samples, where $\boldsymbol{\Theta} = (\boldsymbol{\theta}_1, \ldots, \boldsymbol{\theta}_N) \in \mathcal{X}^N$ are the angular inputs of the parametrized quantum circuit, and $\boldsymbol{\tilde{E}} = (\tilde{E}_1, \ldots, \tilde{E}_N)^\top \in \mathbb{R}^N$ are the energies measured for the corresponding angular parameters $\boldsymbol{\theta}_i\in \boldsymbol{\Theta}$. The posterior of a GP regression model, see~\cref{eq:GP_prior}, is still a GP, i.e., $p(E(\cdot) \mid \boldsymbol{\Theta}, \boldsymbol{\tilde{E}})=\operatorname{GP}\left(E(\cdot) ; \mu_{\boldsymbol{\Theta}}(\cdot), s_{\boldsymbol{\Theta}}(\cdot, \cdot)\right)$. Thus, in Ref.~\cite{nicoli2023physicsinformed}, it was shown that for arbitrary $M$ test points $\boldsymbol{\Theta}^{\prime}=\left(\boldsymbol{\theta}_1^{\prime}, \ldots, \boldsymbol{\theta}^{\prime}{ }_M\right) \in \mathcal{X}^M$, the posterior, or predictive distribution of the function values $\boldsymbol{E}^{\prime}=\left(E\left(\boldsymbol{\theta}_1^{\prime}\right), \ldots, E\left(\boldsymbol{\theta}_M^{\prime}\right)\right)^{\top} \in \mathbb{R}^M$, is an $M$-dimensional Gaussian,
\begin{equation}\label{eq:gp_update}
\begin{gathered}
p\left(\boldsymbol{E}^{\prime} \mid \boldsymbol{\Theta}, \boldsymbol{\tilde{E}}\right)=\mathcal{N}_M\left(\boldsymbol{E}^{\prime} ; \boldsymbol{\mu}_{\boldsymbol{\Theta}}^{\prime}, \boldsymbol{S}_{\boldsymbol{\Theta}}^{\prime}\right),
\end{gathered}
\end{equation}
 with mean and covariance analytically given by
\begin{equation}
\begin{gathered}
\boldsymbol{\mu}_{\boldsymbol{\Theta}}^{\prime}=\boldsymbol{K}^{\prime \top}\left(\boldsymbol{K}+\sigma^2 \boldsymbol{I}_N\right)^{-1} \boldsymbol{\tilde{E}}, \quad \boldsymbol{S}_{\boldsymbol{\Theta}}^{\prime}=\boldsymbol{K}^{\prime \prime}-\boldsymbol{K}^{\prime \top}\left(\boldsymbol{K}+\sigma^2 \boldsymbol{I}_N\right)^{-1} \boldsymbol{K}^{\prime} .
\end{gathered}
\end{equation}
The notation with a prime refers to new, unseen points with the corresponding energies predicted by the GP. Here, $\boldsymbol{K}=k(\boldsymbol{\Theta}, \boldsymbol{\Theta}) \in \mathbb{R}^{N \times N}, \boldsymbol{K}^{\prime}=k\left(\boldsymbol{\Theta}, \boldsymbol{\Theta}^{\prime}\right) \in \mathbb{R}^{N \times M}$, and $\boldsymbol{K}^{\prime \prime}=k\left(\boldsymbol{\Theta}^{\prime}, \boldsymbol{\Theta}^{\prime}\right) \in \mathbb{R}^{M \times M}$ are the train, train-test, and test kernel matrices, respectively, where $k\left(\boldsymbol{\Theta}, \boldsymbol{\Theta}^{\prime}\right)$ denotes the kernel matrix evaluated at each column of $\boldsymbol{\Theta}$ and $\boldsymbol{\Theta}^{\prime}$, such that $\left(k\left(\boldsymbol{\Theta}, \boldsymbol{\Theta}^{\prime}\right)\right)_{n, m}=k\left(\boldsymbol{\theta}_n, \boldsymbol{\theta'}_m\right)$. Moreover, $\boldsymbol{I}_N \in \mathbb{R}^{N \times N}$ denotes the identity matrix. The subscript $\boldsymbol{\Theta}$ in~\cref{eq:gp_update} denotes which input points were used to train the GP. Note that the predictive mean and covariance are given in their vector and matrix form, respectively. The predictive mean can be computed as a function of one arbitrary test point $\boldsymbol{\theta}^{\prime}$, i.e., $\mu_{\boldsymbol{\Theta}}(\boldsymbol{\theta}^\prime) = \boldsymbol{K}^{\prime \top}\left(\boldsymbol{K}+\sigma^2 \boldsymbol{I}_N\right)^{-1} \boldsymbol{\tilde{E}}$ with $\boldsymbol{K}^{\prime}=k\left(\boldsymbol{\Theta}, \boldsymbol{\theta}^{\prime}\right) \in \mathbb{R}^{N}$. The computation of a single predictive covariance $s_{\boldsymbol{\Theta}}(\cdot, \cdot)$ with $ \boldsymbol{S}_{\boldsymbol{\Theta}}^{\prime} =s_{\boldsymbol{\Theta}}\left(\boldsymbol{\Theta}^{\prime}, \boldsymbol{\Theta}^{\prime}\right)\in \mathbb{R}^{M\times M}$ follows analogously. For more details on how to compute the predictive mean and variance, we refer to Ref.~\cite{nicoli2023physicsinformed}.

As the number of measurements increases, the knowledge gained by the GP during optimization can be used not only for regression but also to choose the next points to observe. This is achieved by combining GPR and BO and solving the following maximization problem:
$$
\max_{\boldsymbol{\Theta'}}{a_{\boldsymbol{\Theta}^{(t-1)}}(\boldsymbol{\Theta'})}\, ,
$$
where $a_{\boldsymbol{\Theta}^{(t-1)}}(\cdot)$ is the so-called \textit{acquisition function}.  
This function uses a GP model trained on previously observed data pairs $\{\boldsymbol{\Theta}^{(t-1)},\boldsymbol{\tilde{E}}\}$, collected over the prior $(t-1)$ iterations, to assess how promising a set of new input points $\boldsymbol{\Theta'}$ may be. The points that maximize the acquisition function are likely to be the most informative for the next iteration.

\section{Method}
\label{sec:ProposedMethod}

In~\cref{sec:vqe_kernel}, we leverage the functional form of the VQE objective~\cite{nakanishi20} to construct a \textit{physics-informed kernel} suited for GPR~\cite{nicoli2023physicsinformed}. In~\cref{sec:EMICoRe_acq_func}, we introduce a novel type of acquisition function called \textit{Expected Maximum Improvement over Confident Regions} (EMICoRe)~\cite{nicoli2023physicsinformed}. In~\cref{sec:nft_emicore}, we combine these concepts and present the \textit{NFT-with-EMICoRe algorithm}, called \textit{EMICoRe algorithm} for short~\cite{nicoli2023physicsinformed}. This algorithm demonstrates how the NFT baseline algorithm can be improved to achieve a more flexible and noise-resilient algorithm when combined with BO and GPR. 

\subsection{VQE Kernel}\label{sec:vqe_kernel}

A GP needs to be equipped with a kernel (covariance) function. While there are  many types of such functions, one popular example is the Gaussian-RBF kernel~\cite{rasmussen}. This kernel encodes the assumption that input vectors that are close in the input space exhibit stronger correlations compared to those that are distant, with the correlation strength decaying exponentially as a function of the distance. While this may be a reasonable assumption, an exponential decay in correlation, as seen in the Gaussian-RBF kernel, may not accurately reflect the underlying features of every problem. For this reason, the choice of kernel in a GPR task can significantly affect performance. In Ref.~\cite{nicoli2023physicsinformed} the authors proposed the so-called VQE kernel:
\begin{equation}
k^{\mathrm{VQE}}\left(\boldsymbol{\theta}, \boldsymbol{\theta}^{\prime}\right)=\sigma_0^2 \prod_{d=1}^D\left(\frac{\gamma^2+\cos \left(\theta_d-\theta_d^{\prime}\right)}{1+\gamma^2}\right),
\end{equation}
where $D$ is the number of angular parameters of the parametrized quantum circuit, $\sigma_0$ corresponds to the prior variance, and $\gamma^2 \geq 1$ controls the smoothness of the kernel. Both $\sigma_0$ and $\gamma^2$ are hyperparameters of the kernel.

In~\cite{nicoli2023physicsinformed}, it was shown that the VQE kernel can be decomposed as
\begin{equation}\label{eq:kernel_decomposition}
    k^{\mathrm{VQE}}\left(\boldsymbol{\theta}, \boldsymbol{\theta}^{\prime}\right) = \boldsymbol{\phi}(\boldsymbol{\theta})^\top \boldsymbol{\phi}(\boldsymbol{\theta}^\prime), \quad \textrm{with}  \quad \boldsymbol{\phi}(\boldsymbol{\theta})=\sigma_0\left(1+\gamma^2\right)^{-D / 2} \left[\otimes_{d=1}^D\left(\gamma, \cos \theta_d, \sin \theta_d\right)^{\top}\right],
\end{equation}
where the feature maps $\boldsymbol{\phi}(\boldsymbol{\theta})$ fulfill the same VQE functional form as derived in~\cref{eq:VQE_objective}. Since a GP regression model is only defined in terms of inner products of the feature maps, one just requires access to the kernel form without explicitly specifying $\boldsymbol{\phi}$ ---a procedure known as \textit{kernel trick}~\cite{rasmussen}.

By using the VQE kernel instead of, for example, the Gaussian-RBF kernel, the GP is ensured to sample functions that align with the required functional form from~\cref{eq:VQE_objective}. 
Consequently, as described in~\cref{eq:VQE_objective_one_parameter}, when optimizing over a one-dimensional subspace---i.e., varying only one parameter at a time---three noiseless observed points are sufficient to uniquely identify the target function~\cite{nakanishi20} across the entire subspace. However, in the presence of measurement shot noise, the target function in the subspace can only be identified up to some statistical uncertainty. Therefore, the GP provides not only energy estimates but also their corresponding uncertainties for new measurements in the subspace. 

Furthermore, it has been shown that the decomposition of the VQE kernel in~\cref{eq:kernel_decomposition} can be generalized so that each entry of $\boldsymbol{\theta}$ may parametrize multiple gates at once~\cite{nicoli2023physicsinformed}. This property becomes crucial if the Hamiltonian is invariant under specific symmetries, leading to multiple parameters being identical and requiring simultaneous updates.

\subsection{Expected Maximum Improvement over Confident Regions}\label{sec:EMICoRe_acq_func}

To perform BO in combination with GPR, one should pick a suitable acquisition function to identify the most promising points to measure at each optimization step.   
A novel acquisition function termed \textit{Expected Maximum Improvement over Confident Regions} (EMICoRe) was introduced in the context of GPR for VQEs~\cite{nicoli2023physicsinformed}. This acquisition function relies on the idea of \textit{Confident Region} (CoRe) mathematically defined as
\begin{equation}
\mathcal{Z}_{\boldsymbol{\Theta}}=\left\{\boldsymbol{\theta} \in \mathcal{X} ; s_{\boldsymbol{\Theta}}(\boldsymbol{\theta}, \boldsymbol{\theta}) \leq \kappa^2\right\}.
\end{equation}
The CoRe describes all points for which the \textit{predictive} covariance $s_{\boldsymbol{\Theta}}(\cdot, \cdot)$ is smaller than or equal to a threshold $\kappa^2$, which is a hyperparameter of the model. For sufficiently small $\kappa$, an appropriate kernel function, and a weak prior---i.e., enough previously observed points---every point in the CoRe can be regarded as ``already observed'' due to the sufficiently small uncertainty of the GP. Leveraging the concept of CoRe, Nicoli et al.~\cite{nicoli2023physicsinformed} proposed the following acquisition function:
\begin{equation}\label{eq:acq_emicore}
a^{\mathrm{EMICoRe}}\left(\boldsymbol{\Theta}^{\prime}\right)=\frac{1}{M}\left\langle\max \left(0, \min _{\boldsymbol{\theta} \in \mathcal{Z}_{\boldsymbol{\Theta}}} E(\boldsymbol{\theta})-\min _{\boldsymbol{\theta} \in \mathcal{Z}_{\widetilde{\boldsymbol{\Theta}}}} E(\boldsymbol{\theta})\right)\right\rangle_{p(E(\cdot) \mid \boldsymbol{\Theta}, \boldsymbol{\tilde{E}})},
\end{equation}
where $\widetilde{\boldsymbol{\Theta}}=\left(\boldsymbol{\Theta}, \boldsymbol{\Theta}^{\prime}\right) \in \mathcal{X}^{N+M}$ denotes the augmented training set with the new input points $\boldsymbol{\Theta}^{\prime} \in \mathcal{X}^M$ and the previously observed ones $\boldsymbol{\Theta}\in \mathcal{X}^N$.\\
The term $\min _{\boldsymbol{\theta} \in \mathcal{Z}_{\boldsymbol{\Theta}}} E(\boldsymbol{\theta})$ in~\cref{eq:acq_emicore} corresponds to the estimated minimum of the CoRe, identified by all previously observed points $\boldsymbol{\Theta}$. This value is drawn from the GP's posterior distribution and identifies the estimated minimum energy for any choice of circuit parameters previously observed. The other term in~\cref{eq:acq_emicore}, $\min _{\boldsymbol{\theta} \in \mathcal{Z}_{\widetilde{\boldsymbol{\Theta}}}} E(\boldsymbol{\theta})$, is the estimated minimum  over the augmented training set $\widetilde{\boldsymbol{\Theta}}=\left(\boldsymbol{\Theta}, \boldsymbol{\Theta}^{\prime}\right)$. Note that within this augmented set, the new input points $\boldsymbol{\Theta}^{\prime} \in \mathcal{X}^M$ can be treated as if they had already been observed. This is because, if a new input point lies within the CoRe, the GP's uncertainty is below a given threshold, allowing the reasonable assumption that the GP is \textit{highly confident} close to that point. Consequently, this point can be implicitly considered \textit{known}, i.e., as if it has been observed, despite no measurement being performed on the quantum computer.

%\subsubsection{EMICoRe}
\subsubsection{The EMICoRe algorithm}
\label{sec:nft_emicore}

To address the challenges posed by noisy measurements on quantum hardware, Nicoli et al.~\cite{nicoli2023physicsinformed} proposed to combine NFT with GP regression and BO, thus merging all the tools previously introduced in~\cref{sec:Pre.GPandBO,sec:vqe_kernel,sec:EMICoRe_acq_func}.

The proposed EMICoRe algorithm builds upon the structure of the NFT algorithm, incorporating two key modifications: 
\begin{enumerate}
\item \textbf{Learnable shifts}: Instead of taking a fixed shift $\alpha$ to identify two new (equidistant) candidate points, see~\cref{eq:shifts}, the EMICoRe algorithm uses the EMICoRe acquisition function from~\cref{sec:EMICoRe_acq_func} to identify the most \textit{informative} points to measure next. 
\item \textbf{Re-use of previous measurements}: At each iteration, the minimum on the one-dimensional subspace is estimated using Least Square Regression with the posterior mean of the GP. This implies that as more measurements are made on the quantum computer and incorporated into the surrogate model, the GP will provide increasingly accurate estimates.
\end{enumerate}
These modifications distinguish the proposed algorithm from NFT, where the minimum estimation on the subspace depends solely on the previous best point and the \textit{fixed} shifts for the new measurements. 
Both modifications aim to make EMICoRe a more flexible and resilient algorithm in the presence of quantum noise. The EMICoRe algorithm starts with $T_{\textrm{NFT}}$ steps of the baseline NFT algorithm to gather sufficient measurements for instantiating an informative GP that can make predictions. After this initial phase, the EMICoRe algorithm begins learning the shifts at each step, using the previous measurements and storing the new measurements for future predictions. For a detailed description of the EMICoRe algorithm and its subroutine, we refer to the original paper~\cite{nicoli2023physicsinformed}. 

\begin{figure}[tp]
    \centering
    \includegraphics[width=0.9\textwidth]{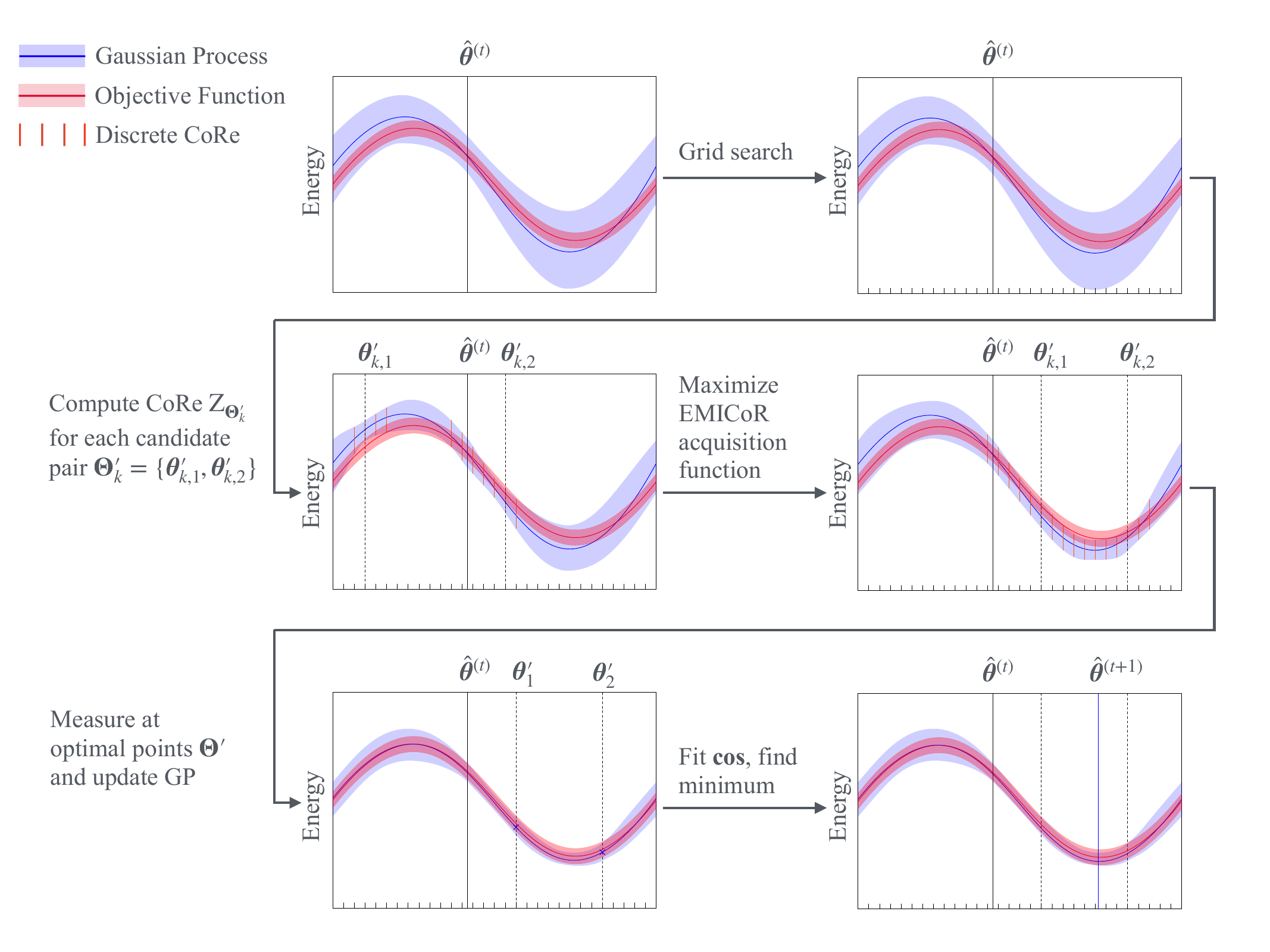}
    \caption{Visualization of the EMICoRe algorithm. We refer to the main text for more details.}
    \label{fig:visualization_EMICoRe}
\end{figure}

\Cref{fig:visualization_EMICoRe} visualizes the algorithmic procedure of the full EMICoRe algorithm. The objective function $E^\ast(\boldsymbol{\theta})$ (true energy) is represented by the red lines, while the red shaded area indicates the standard deviation of a measurement on the quantum computer. The blue lines indicate the predictive mean of the GP $\mu_{\boldsymbol{\Theta}}(\cdot)$, while the shaded area denotes its predictive uncertainty $\sqrt{s_{\boldsymbol{\Theta}}(\cdot,\cdot)}$, i.e., one standard deviation. The figure should be read from top left to bottom right following the black arrow. 

The top left plot shows the optimization in the direction $\theta_d$ for $d \in \{1, \ldots, D\}$\footnote{Note the direction to be optimized is chosen sequentially until $d=D$, after which the algorithm restarts from $d=0$.}. The quantity ${\boldsymbol{\hat{\theta}}}^{(t)}$ denotes the set of parameters which give the lowest energy $E({\boldsymbol{\hat{\theta}}}^{(t)})$ from the GP after the $t$-th optimization step. Note that at this point, the GP has access to a dataset of previous measurements $\{\boldsymbol{\Theta}^{(t)}, \boldsymbol{\tilde{E}}^{(t)}\}$. This explains why the blue line, along with the corresponding uncertainty, provides a good prediction of the red curve along direction $d$, even before any optimization has been performed. 

The EMICoRe subroutine is now responsible for finding the promising points to be measured next, i.e., $\boldsymbol{\Theta}'$ in~\cref{eq:shifts} with the shifts $\alpha$ being different for $\theta_1'$ and $\theta_2'$. First, the subroutine divides the subspace into a grid of equally spaced points, as represented by the grid lines at the bottom of the second plot. Second, the subroutine prepares all possible combinations of candidate pairs $\boldsymbol{\Theta}^\prime_k = \{\boldsymbol{\theta}^\prime_{k,2}, \boldsymbol{\theta}^\prime_{k,2} \}$. These points are added to the \textit{base-GP} in pairs, allowing a discretized CoRe to be computed for each pair (orange vertical lines). The subroutine then computes the acquisition function for each candidate pair $\boldsymbol{\Theta}^\prime_k$ using the respective CoRe. The pair that maximizes the acquisition function, as described in~\cref{eq:acq_emicore}, is returned as $\boldsymbol{\Theta}'$. According to the algorithm, this pair represents the points that are the most promising to be observed next. The fifth plot displays the measurements on the quantum computer at the suggested points, indicated by the blue crosses. The GP is finally updated using the new measurements---note the change of the blue curve and the shaded area in the plot. Finally, the GP's posterior mean is computed for three equidistant points on the one-dimensional subspace. These points are then used to fit a cosine function, according to the functional form of the subspace given by~\cref{eq:VQE_objective_one_parameter}. This fitted cosine function is then used to obtain the analytic minimum of the one-dimensional subspace. 

We recall that, due to the special form of the VQE-kernel and the design of the EMICoRe algorithm, the function sampled from the GP---i.e, the posterior mean---follows the same functional form as in~\cref{eq:VQE_objective}. Therefore, fitting a sinusoidal function to the GP's posterior is a valid assumption.

\section{Numerical Experiments}
\label{sec:experiments}

In the following, we perform numerical experiments to demonstrate the performance of EMICoRe in the presence of simulated hardware noise.
Our experiments focus on the quantum Heisenberg Hamiltonian, a typical Hamiltonian for benchmarking the performance of VQEs,
\begin{equation}
H=-\left[\sum_{j=1}^{Q-1}\left(J_X X_j X_{j+1}+J_Y Y_j Y_{j+1}+J_Z Z_j Z_{j+1}\right)+\sum_{j=1}^Q\left(h_X X_j+h_Y Y_j+h_Z Z_j\right)\right],
\end{equation}
where the index of the Pauli matrices $\{X_j,Y_j,Z_j\}$ refers to the qubit that the Pauli gate is applied to, and open boundary conditions are considered. In this work, we choose the couplings to be \textit{at the critical point} with $(J_X, J_Y, J_Z) = (-1,0,0)$ and $(h_X, h_Y, h_Z) = (0,0,-1)$. The resulting Hamiltonian is also known as the Ising Hamiltonian \textit{at criticality}.

In order to assess the performance of the VQE during optimization, we mainly consider two metrics: \textit{energy} and \textit{fidelity}.
The former is the lowest energy reached by the VQE during the optimization process, i.e., the expectation value of the Hamiltonian with respect to the variational state. The latter measures the overlap between the variational state and the true ground state obtained with exact diagonalization. 
We note that the true ground-state energy $\underline{E}^\ast =\langle\psi_\mathrm{GS}|H|\psi_\mathrm{GS}\rangle$, along with with the corresponding true ground-state wave function $|\psi_\mathrm{GS}\rangle$, can be obtained in our case, as the size of the systems we address is sufficiently small. This is important as it provides access to a ground truth for benchmarking different algorithms, such as NFT versus EMICoRe. We refer to~\cite{nicoli2023physicsinformed} for more details on the metrics we used.  
To test the stability of the optimization with respect to initialization, we initialize  several independent runs with different seeds. Specifically, we use $50$ runs for the experiments without hardware noise and $10$ runs for the experiments with hardware noise, see~\cref{sec:exp_hn} for more details. In all plots, we report the median (50th percentile) as a solid line, with the 25th and 75th percentiles as the lower and upper bounds, respectively, represented by shaded regions around the solid line. The results are plotted as a function of the total number of measurements performed on the quantum computer. 

For all simulations, we use an $L$-layered and fully connected \texttt{Efficient SU(2)} circuit with $D = 2(L + 1)\cdot Q$ total angular parameters. The classical simulations of the quantum hardware are performed with the Qiskit~\cite{qiskit2024} library, while the Python implementation of the EMICoRe algorithm has been adapted from the original code available on GitHub~\cite{emicore_GH_2023}.

All the parameters for EMICoRe are identical to those chosen for the simulation in Fig.\ 7 of the original paper~\cite{nicoli2023physicsinformed}. As noted in \cref{footnote}, we used $\alpha=\pi/3$ as the shift for NFT. Moreover, in contrast to the original NFT paper~\cite{nakanishi20}, we did not include a stabilization step, as our results indicated that its use often led to poorer performance. Further investigation of this phenomenon is left to future work. 

\begin{figure}
    \centering
    \includegraphics[width=0.35\textwidth]{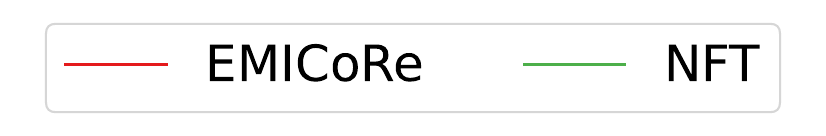}
    \vspace{1em} % Adds vertical space between subfigures
    \begin{subfigure}[b]{0.85\textwidth}
        \centering
        \includegraphics[width=\textwidth]{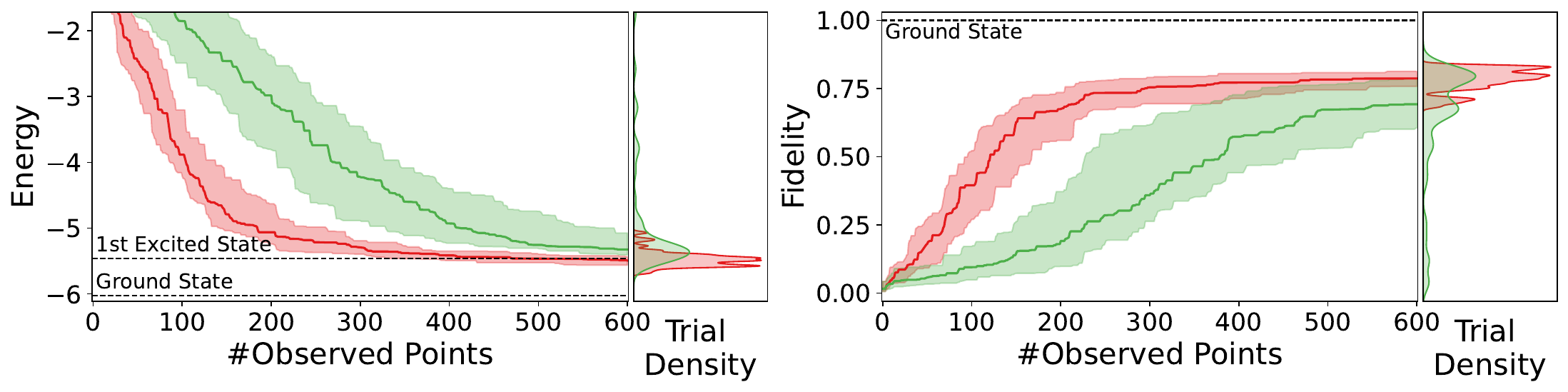}
        \caption{Classical simulation of noisy quantum hardware  without error mitigation.}
        \label{fig:5-3-ising-noise}
    \end{subfigure}
    \begin{subfigure}[b]{0.85\textwidth}
        \centering
        \includegraphics[width=\textwidth]{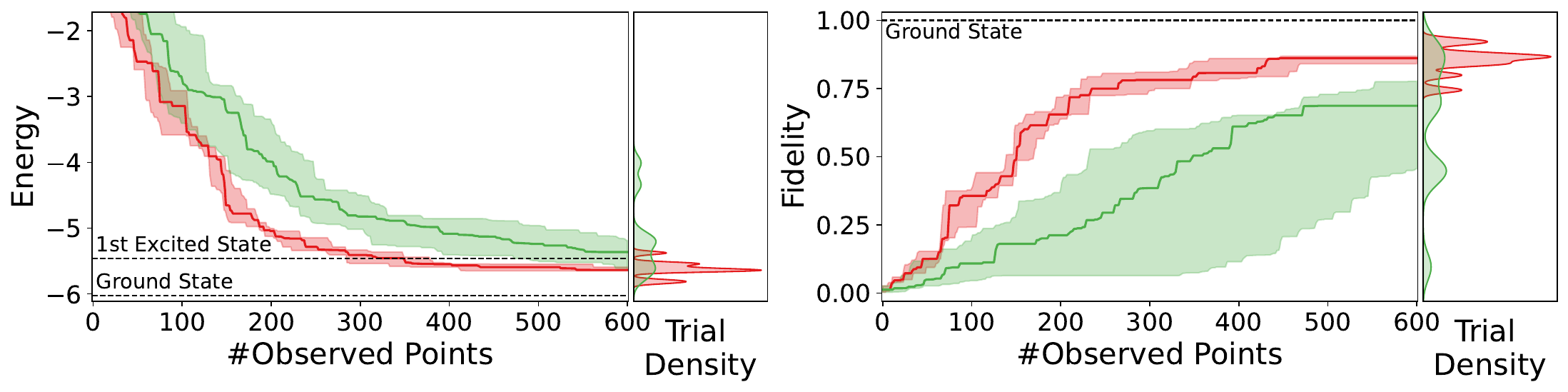}
        \caption{Classical simulation of noisy quantum hardware  with TREX error mitigation.}
        \label{fig:5-3-ising-noise-trex}
    \end{subfigure}
    \begin{subfigure}[b]{0.85\textwidth}
        \centering
    \includegraphics[width=\textwidth]{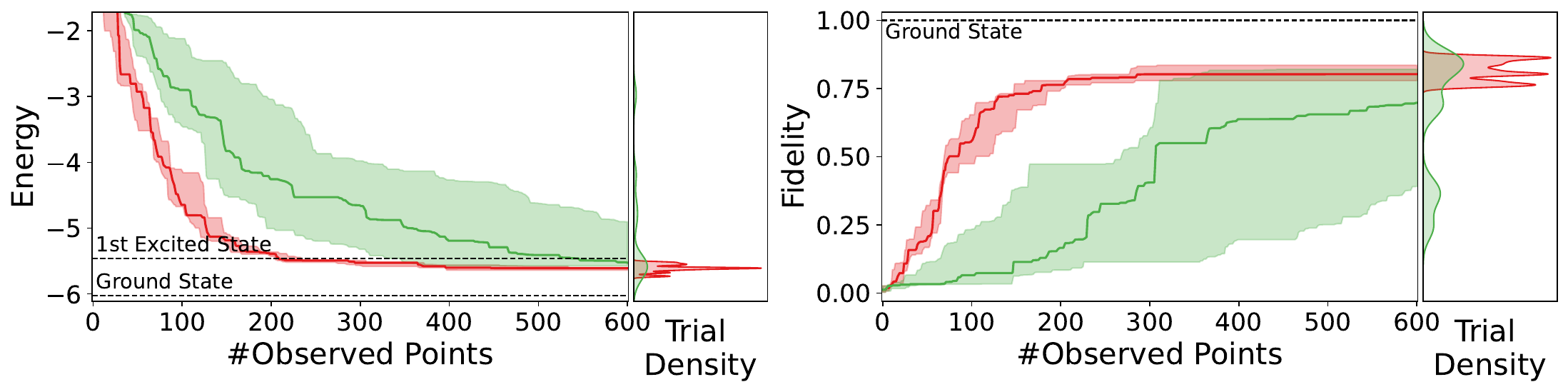}
        \caption{Classical simulation of noisy quantum hardware with ZNE error mitigation.}
        \label{fig:app_5-3-ising-noise-zne}
       \end{subfigure}
    \caption{Energy (left) and fidelity (right) for NFT (green) and EMICoRe (red). We show results for the critical Ising Hamiltonian using $Q = 5$ qubits and $L = 3$ layers, for an Efficient SU(2) quantum circuit,  and $N_{\textrm{shots}} = 1024$ measurement shots. The top row displays runs with simulated hardware noise and no error mitigation, while the second and third rows present results with TREX and ZNE error mitigation schemes, respectively. The median (solid line) and the $25^\mathrm{th}$ and $75^\mathrm{th}$ percentiles (colored regions surrounding the median) are obtained with $50$ independent seeded runs for experiments in~\cref{fig:5-3-ising-noise} and with $10$ runs for~\cref{fig:5-3-ising-noise-trex,fig:app_5-3-ising-noise-zne}. A density of the trial distribution at the end of the optimization is shown right next to each plot.}
    \label{fig:results}
\end{figure}
\subsection{Hardware Noise Analysis of EMICoRe}
\label{sec:exp_hn}
Previous work~\cite{nicoli2023physicsinformed} analyzed EMICoRe's performance in the presence of measurement shot noise, showing that EMICoRe outperforms NFT more significantly at higher noise levels, i.e., with fewer shots. In these proceedings, we extend the analysis to include hardware noise as well. 

In the following experiments, the setup remains constant: we consider a $5$-qubit system with $3$-layers of the \texttt{EfficientSU2} circuit and the Ising Hamiltonian with open boundary conditions. All simulations are conducted with a total of $600$ measurements~\footnote{Note that two measurements are performed at each step, meaning that $600$ measurements correspond to $300$ optimization steps.}, using $N=1024$ shots per measurement. 

In~\cref{fig:5-3-ising-noise}, we conduct experiments with simulated hardware noise and no error mitigation. In~\cref{fig:5-3-ising-noise-trex} and~\cref{fig:app_5-3-ising-noise-zne}, we apply \textit{twisted readout error extinction} (TREX)~\cite{TREX} and \textit{zero noise extrapolation} (ZNE)~\cite{Temme_2017, Li_2017} as error mitigation strategies, respectively. Hardware noise is simulated with the \texttt{Fake5QV1} backend. The results in~\cref{fig:5-3-ising-noise} are obtained by aggregating $50$ independent runs, whereas~\cref{fig:5-3-ising-noise-trex,fig:app_5-3-ising-noise-zne} include only $10$ runs, as the feature was deprecated before further data could be collected. \cref{tab:simulation_error} summarizes the results from~\cref{fig:results}, reporting the mean and standard deviation computed over multiple seeded trials for simulations with and without error mitigation.

When hardware noise is taken into account, EMICoRe significantly outperforms NFT, as shown in~\cref{fig:5-3-ising-noise}, demonstrating faster and more reliable convergence. For instance, the trial density on the right-hand-side is more peaked around the median. These empirical results indicate that NFT is considerably more affected by hardware noise compared to EMICoRe. While neither algorithm reaches the ground-state energy, EMICoRe achieves a fidelity of nearly $80\%$.
With error mitigation, EMICoRe outperforms NFT in both energy minimization and fidelity, demonstrating greater stability across different initializations and achieving lower energies as well as higher fidelities. 

\begin{table}[tp]
\centering
\caption[Mean and std. of Ising with noise]{Mean and standard deviation of the energy and fidelity at the final optimization step for all simulations presented in~\cref{fig:results}. For the energy, lower values are better; for the fidelity, higher values are better.}
\label{tab:simulation_error}
\begin{tabular}{||l||l|l||l|l||} 
\hhline{|t:=:t:==:t:==:t|}
\multirow{2}{*}{} & \multicolumn{2}{l||}{EMICoRe}                               & \multicolumn{2}{l||}{NFT}                                    \\ 
\cline{2-5}
                                                                                & Energy $E^\ast({\boldsymbol{\hat{\theta}}})$ & Fidelity      & Energy $E^\ast({\boldsymbol{\hat{\theta}}})$ & Fidelity       \\ 
\hhline{|:=::==::==:|}
% no noise                                                                                 & $(-5.817 \pm 0.164)$                         & $(0.863 \pm 0.165)$ & $(-5.814 \pm 0.167)$                         & $(0.837 \pm 0.187)$  \\ 
% \hline
noise                                                                                    & $(-5.472 \pm 0.119)$                         & $(0.782 \pm 0.04)$  & $(-4.972 \pm 0.821)$                         & $(0.641 \pm 0.214)$  \\ 
\hline
noise, TREX                                                                           & $(-5.624 \pm 0.122)$                         & $(0.854 \pm 0.051)$ & $(-5.173 \pm 0.543)$                         & $(0.622 \pm 0.245)$  \\ 
\hline
noise, ZNE                                                                            & $(-5.611 \pm 0.059)$                         & $(0.814 \pm 0.037)$ & $(-5.03 \pm 0.841)$                          & $(0.653 \pm 0.226)$  \\
\hhline{|b:=:b:==:b:==:b|}
\end{tabular}
\end{table}

\section{Conclusion}
\label{ref:conclusion}

In these proceedings, we presented a novel machine-learning-enhanced optimization approach for VQE algorithms to find the ground state of the quantum Ising Hamiltonian in the presence of simulated hardware noise. This algorithm, called EMICoRe~\cite{nicoli2023physicsinformed}, combines Gaussian Process Regression for noise-resilient energy estimation and Bayesian Optimization to predict new candidate points for measurement at each step of the optimization.

Our numerical experiments demonstrated that EMICoRe outperforms the state-of-the-art baseline, the NFT algorithm~\cite{nakanishi20}, in the presence of simulated hardware noise. EMICoRe converges to lower energies than NFT and shows greater robustness across different initializations. This robustness underscores the potential of the proposed method for future applications to more complex quantum systems and under the challenging conditions of real NISQ devices.

\section*{Acknowledgments}
\noindent
The authors thank Christopher J.\ Anders and Shinichi Nakajima for insightful discussions. This project was supported by the Deutsche Forschungsgemeinschaft (DFG, German Research Foundation) as part of the CRC 1639 NuMeriQS -- project no. 511713970.

\bibliographystyle{JHEP}
\bibliography{refs}
\end{document}